\documentclass[12pt,aasmacros]{spieman}  
\usepackage{amsmath,amsfonts,amssymb}
\usepackage{graphicx}
\usepackage{setspace}
\usepackage{tocloft}
\usepackage{xcolor}
\usepackage{xspace}
\usepackage{lineno}

\title{Testing Station for Fast Screening of Through Silicon Via (TSV)-enabled Application Specific Integrated Circuits (ASICs) for Hard X-ray Imaging Detectors}

\author[1]{Daniel P. Violette$^{\star}$\textsuperscript{$\dagger$}}
\author[1]{Branden Allen}
\author[1]{Jaesub Hong}
\author[2]{Hiromasa Miyasaka}
\author[1]{Jonathan Grindlay}

\affil[1]{Center for Astrophysics $|$ Harvard \& Smithsonian, 60 Garden St, Cambridge, MA 02138, USA}
\affil[2]{California Institute of Technology, Pasadena, CA 91125, USA}

\cftpagenumbersoff{figure}
\cftpagenumbersoff{table} 
\begin{document} 
\maketitle

\begin{abstract}

Application Specific Integrated Circuits (ASICs) are used in space-borne instruments for signal processing and detector readout. The electrical interface of these ASICs to frontend printed circuit boards (PCBs) is commonly accomplished with wire bonds. Through Silicon Via (TSV) technology has been proposed as an alternative interconnect technique that will reduce assembly complexity of ASIC packaging by replacing wire bonding with flip-chip bonding. TSV technology is advantageous in large detector arrays where TSVs enable close detector tiling on all sides. Wafer-level probe card testing of TSV ASICs is frustrated by solder balls introduced onto the ASIC surface for flip-chip bonding that hamper alignment. Therefore, we developed the ASIC Test Stand (ATS) to enable rapid screening and characterization of individual ASIC die. We successfully demonstrated ATS operation on ASICs originally developed for CdZnTe detectors on the Nuclear Spectroscopic and Telescope Array (\textit{NuSTAR}) mission that were later modified with TSVs in a via-last process. We tested both back-side blind-TSVs and front-side through-TSVs, with results from internal test pulser measurements that demonstrate performance equal to or exceeding the  probe card wafer-level testing data. The ATS can easily be expanded or duplicated in order to parallelize ASIC screening for large area imaging detectors of future space programs.
\end{abstract}


{\noindent \footnotesize{$^{\star}$daniel.violette@cfa.harvard.edu}} \\
{\noindent \footnotesize{\textsuperscript{$^\dagger$}NASA FINESST Fellow}}

\section{Introduction \& Motivation}

\begin{figure*}[tbh!]
    \centering
    \includegraphics[width=6.5in]{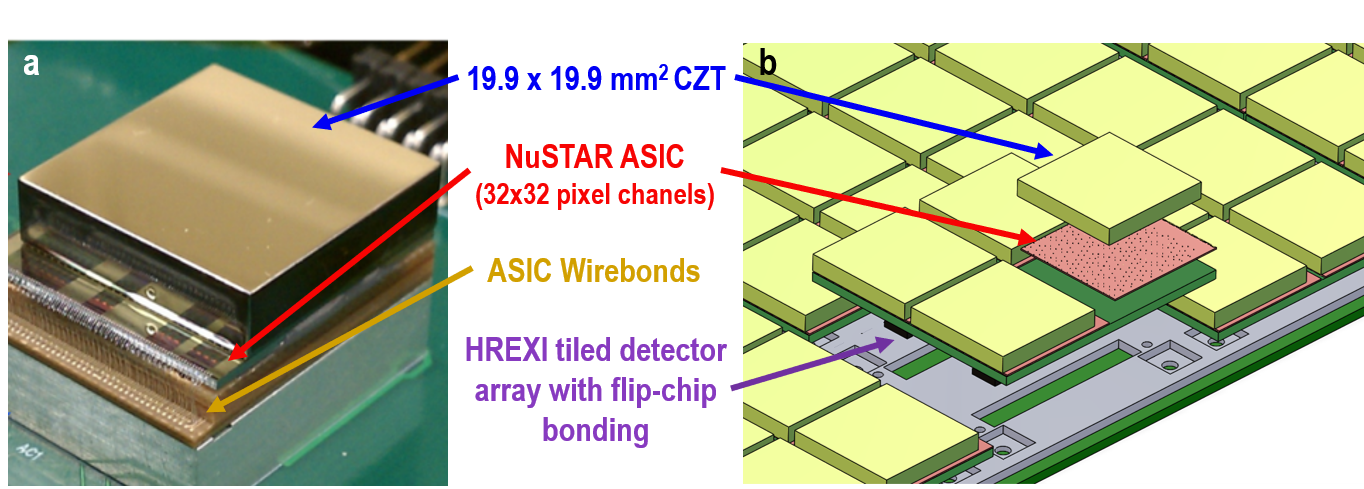}
\caption{(\textbf{\textit{a}}) A single detector crystal unit (DCU) of a CZT crystal bonded with conductive epoxy to a NuASIC. The NuASIC is wire bonded to a frontend PCB to power and control the detector. (\textbf{\textit{b}}) CAD model of the HREXI coded aperture instrument, composed of tiled arrays of 2\,$\times$\,2 DCUs. Flip-chip bonding TSV NuASIC allows for tightly packed detector spacing.}
    \label{f:detector}
\end{figure*}

Wide-field coded aperture hard X-ray telescopes enable the discovery and characterization of the most energetic and transient astrophysical phenomena in the Universe, including Gamma-ray bursts, outbursts from supermassive black holes with jets (blazars), black hole and neutron star binary mergers, and outbursts from black hole and neutron star X-ray binaries. A next-generation instrument under development is the High Resolution Energetic X-ray Imager (HREXI). HREXI is a CdZnTe (CZT) imaging hard X-ray (3-300 keV) detector designed to identify and localize transient astrophysical events. Drawing on heritage technology developed during the \textit{ProtoEXIST} \textit{1} \& \textit{2} balloon-borne X-ray telescope experiments\cite{Hong09,Allen11,Hong13,Hong17}, HREXI is composed of a modular array of 256 closely tiled pixellated 19.9\,$\times$\,19.9~mm$^{2}$, 3~mm thick CZT detectors. Each CZT detector is bonded to an individual Application Specific Integrated Circuit (ASIC) with gold studded conductive-epoxy (Fig.~\ref{f:detector}a). Previously developed and flown on the the \textit{NuSTAR} mission\cite{Harrison13}, the \textit{NuSTAR} ASIC (NuASIC)\cite{Harrison10} readout enables tiling of large detector area assemblies for coded aperture imaging (Fig.~\ref{f:detector}b). While \textit{NuSTAR} operates eight individual detectors (2\,$\times$\,2 for each of the two focusing telescopes), a coded aperture X-ray telescope requires orders of magnitude more detectors to achieve desired sensitivities over a larger field of view. Fabrication, testing and assembly of hundreds of CZT and NuASIC detectors will be labor intensive and every improvement to the integration process will result in large reductions to schedule time and cost. To overcome this challenge, the HREXI program plans to replace NuASIC wire bonds with Through Silicon Vias (TSVs)\cite{Hong21}. In this paper, we detail efforts to streamline screening and testing of TSV NuASICs before bonding to CZT.

Wire bonding is a standard space-qualified technique for connecting integrated circuits with other electronics that are commonly used for detector and sensor integration. Wire bonds are thin gold wires attached to pads on the NuASIC's surface which are each connected to staggered pads on a frontend electronics board beneath.  The use of exposed wire bonds has a number of drawbacks, including vulnerability to damage that complicate the safe integration of large detector planes. Attempting to protect wire bonds through encapsulation will add additional processing and integration time for each detector unit. Furthermore, the frontend board must be larger than the NuASIC to provide room for bonding, creating gaps between individual NuASICs which quickly scales the size of large detector arrays. The necessary gaps between detectors also introduce spatial non-uniformity in the instrument background by allowing X-rays to penetrate the detectors through the side walls of the CZT crystals\cite{Hong13}. Wire bonds may also act as a source of electrical noise pickup during detector operation. Due to these disadvantages, we explore an alternative interconnect solution for large area tiled arrays.

\begin{figure*}[tbh!]
   \centering
   \includegraphics[width=5.0in]{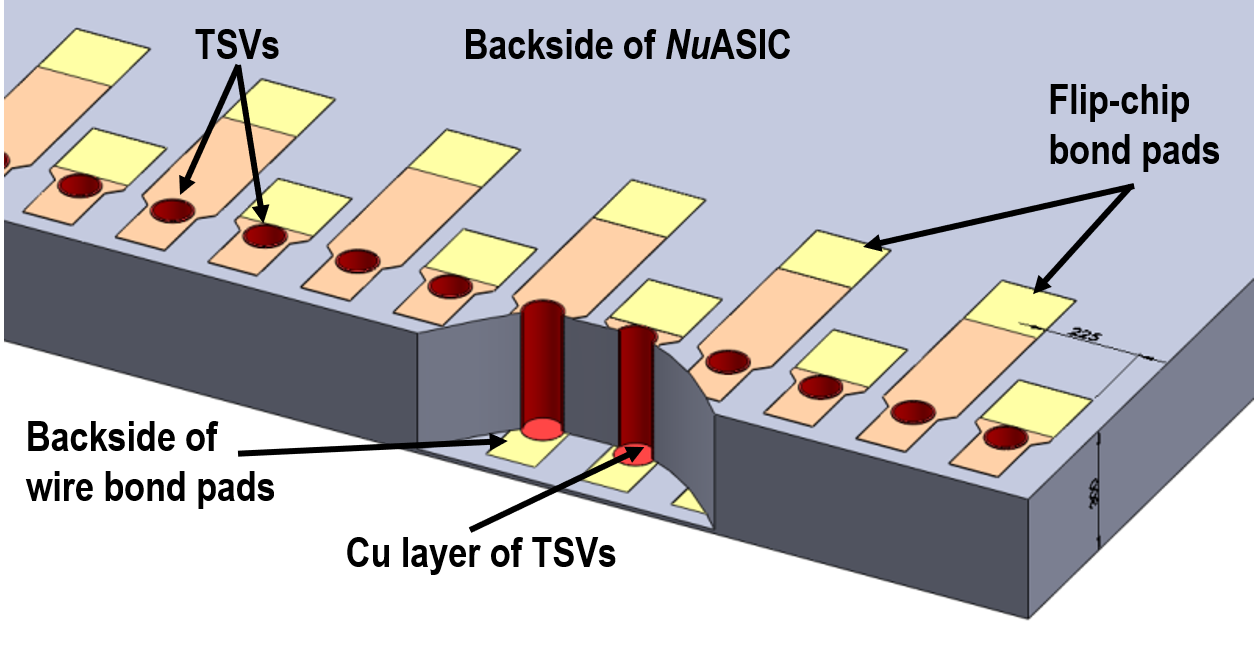}
\caption{A depiction of back-side \textit{blind}-TSVs that connect the single row of wire bond pads on the NuASIC's top surface with a double row of flip-chip bond pads on the NuASIC's bottom surface. The flip-chip bond pads are offset to prevent the attached solder balls from merging together during attachment\cite{Hong17b}.}
    \label{f:tsv}
\end{figure*}

TSVs offer an alternative to wire bonds for connecting each NuASIC to the readout electronics system\cite{Hong21}. TSVs are metallized vias that connect the wire bond pads on the top surface of the NuASIC with a staggered array of pads on the bottom surface (Fig.~\ref{f:tsv})\cite{Hong17b}. These new pads can then be easily flip-chip solder-bonded to a frontend board below the NuASIC, removing the need for wire bonds entirely. This eliminates the handling and tiling concerns introduced by the wire bond contacts. In a ``via-last'' approach, the NuASIC's 8-inch silicon wafers are mounted to a carrier wafer for mechanical stability before thinning, etching and metallization as described in Hong et al.\,(2021)\cite{Hong21}. After TSV implementation, traces and pads are plated to the back side of the NuASIC before lead solder balls are deposited on the NuASIC pads for later flip-chip bonding. Finally, the carrier wafer is detached and the new TSV NuASICs are diced into 49 individual devices. The additional TSV processing will likely reduce the overall yield of NuASICs per wafer. Further testing of the functionality of each TSV NuASIC must therefore be performed prior to detector bonding to a frontend PCB and CZT.

For \textit{NuSTAR}, testing of all NuASICs was performed at the wafer level by carefully aligning a probe card over the 87 (225 $\mu$m pitch) wire bond pads for the power, command, and signal lines of a single NuASIC. After confirming successful NuASIC operation, a diagnostic test was performed with an internal test pulser to probe the functionality and resolution of each of the device's 1024 pixel channels. To test NuASICs on the wafer sequentially required the time-consuming realignment of the test probe card over each device. This testing process was suitable for the \textit{NuSTAR} mission which ultimately selected the eight best candidate NuASIC detectors, but is untenable for wide-field coded aperture detector arrays like HREXI. The difficulties with probe card testing are exacerbated by the addition of solder balls for flip-chip bonding on the back-side of the NuASIC. The rounded surface and uneven heights of solder balls disrupts probe placement and creates uncertainties in alignment. Considering the reduced NuASIC yield from the via-last TSV insertion process, the disadvantages of traditional probe card testing, and the benefits of testing individual TSV NuASICs prior to bonding, we designed and produced the ASIC Test Stand (ATS).


\section{{Nu}ASICs Testing Requirements and ATS Design} \label{s:design}

\begin{figure}[tbh!]
   \centering
\includegraphics[width=5.0in]{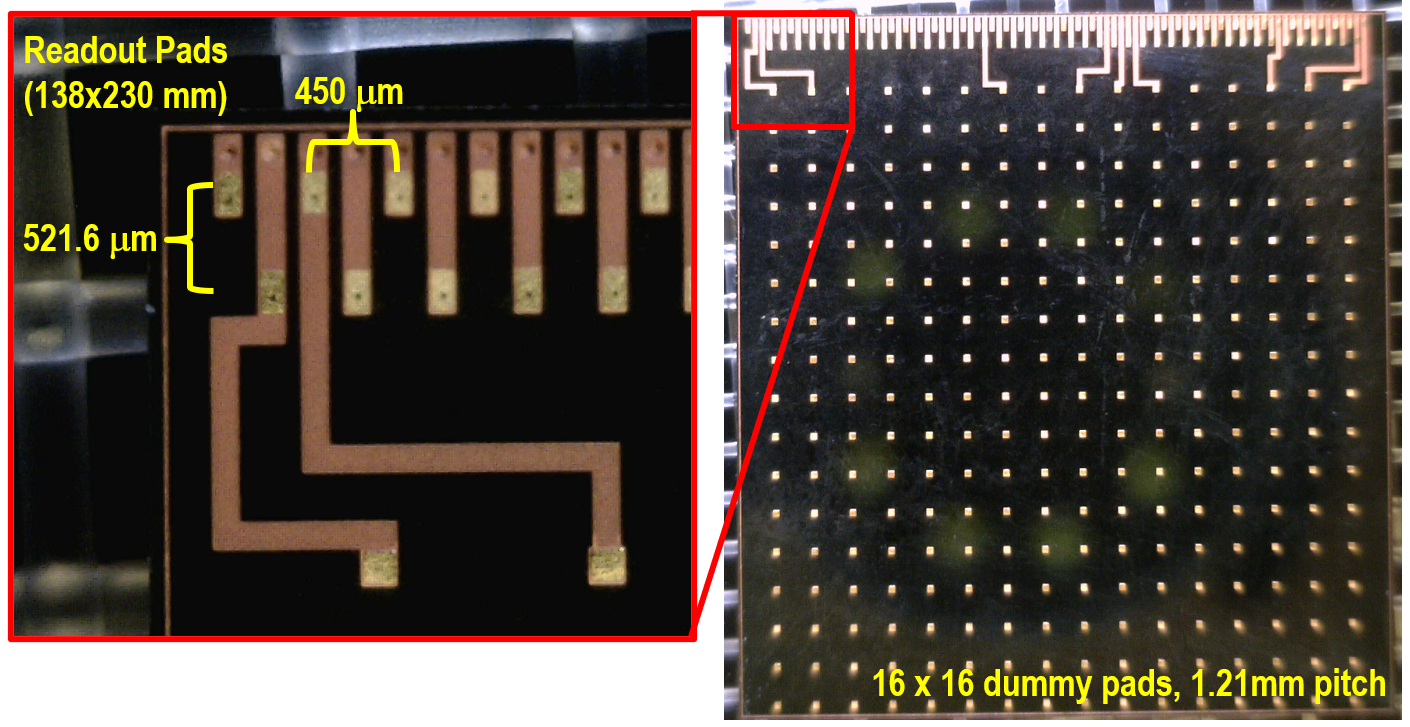}
\caption{Bottom-side view of the TSV NuASIC. 87 power, readout, and commanding traces and pads are arrayed in 2 staggered rows across the top edge of the NuASIC with a 450 $\mu$m pitch between adjacent pads in a row. 16\,$\times$\,16 “dummy pads” (no connections) fill out the remaining area with a 1.209 mm pitch and are used for flip-chip bonding to the readout board.}
    \label{f:asic}
\end{figure}

The ATS is designed to support rapid testing of unattached TSV NuASICs\cite{Violette18}. TSV NuASICs have a single row of 87 wire bond pads for command, data and power on the top surface. Conductive TSVs connect these pads to two staggered rows of traces along an edge on the bottom surface of the NuASIC for flip-chip bonding (Fig.~\ref{f:asic}). These two offset rows of contacts have a 450 $\mu$m pitch (225 $\mu$m trace pitch) and a 138 $\mu$m\,$\times$\,230 $\mu$m pad size. Additionally, a grid of 16\,$\times$\,16 (1.209 mm pitch) 188 $\mu$m\,$\times$\,188 $\mu$m unconnected ``dummy'' pads used in the flip-chip bonding process are distributed across the bottom surface of the NuASIC. Each of the 87 operating lines and 256 dummy pads have attached solder balls. The solder balls settle at different heights (with a $\sim$12~$\mu$m tolerance) on the operational pads and dummy pads due to the difference in pad sizes affecting the solder spread. The variance in solder ball height and tight pitch requirements of the NuASICs necessitates a specialized fixture for rapid testing. Finally, the wafer dicing process introduces die-to-die variance (estimated at 50 $\mu$m) in NuASIC dimensions. Any designed test fixture must be adjustable to account for these NuASIC shape irregularities. 

\begin{figure*}[tbh!]
   \centering
   \includegraphics[width=4.0in]{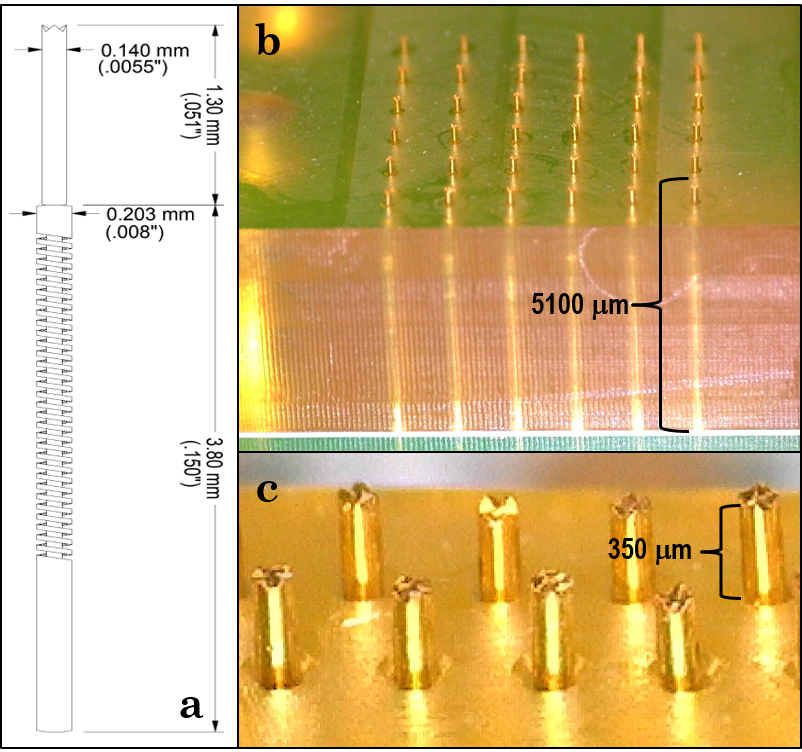}
\caption{(\textbf{\textit{a}}) Diagram of the AlphaTest micro-pogo probe S200 series with crown-tip. (\textbf{\textit{b}}) An array of 6\,$\times$\,6 micro-pogo probes at 1.2 mm pitch encapsulated in the micro-pogo pin test socket (orange) and mounted to test readout board. (\textbf{\textit{c}}) Several micro-pogo probes of the 87 needed for NuASIC readout. These 140 $\mu$m diameter micro-pogo probes can compress up to 350 $\mu$m for full contact.}
    \label{f:pins}
\end{figure*}

\begin{figure*}[tbh!]
   \centering
\includegraphics[width=6.5in]{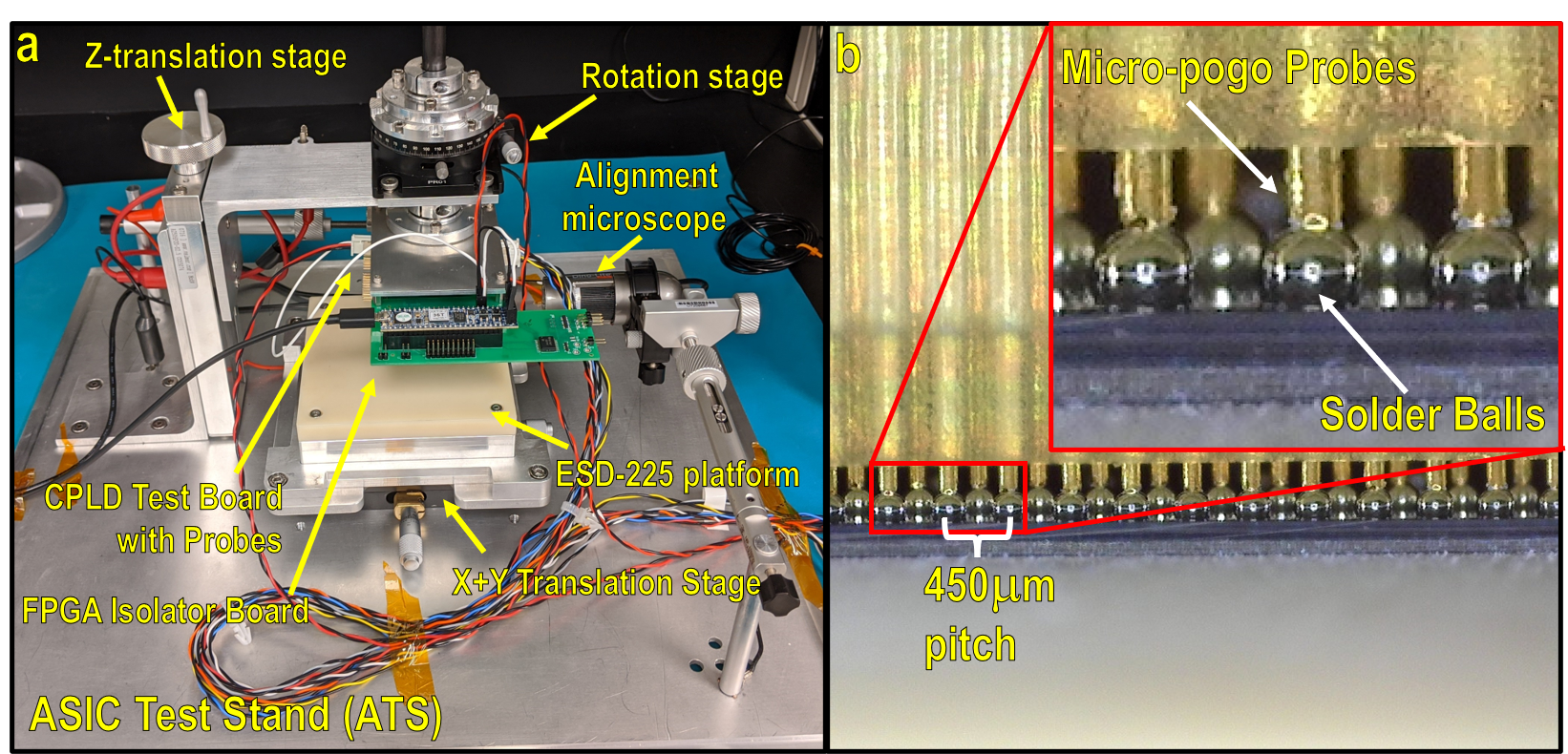}
\caption{(\textbf{\textit{a}}) The ASIC Test Stand (ATS). NuASICs under test are placed on the ESD platform, with alignment controlled by X+Y translation stage and rotation stage. (\textbf{\textit{b}}) Confirmation of {Nu}ASIC contact with test stand micro-pogo probes can be made from edge-on using the ATS's USB microscope stands.}
    \label{f:ats}
\end{figure*}

The ATS utilizes a micro-pogo probe test socket to meet the testing requirements imposed by the TSV NuASIC. Micro-pogo probes are conductive, spring-loaded contacts that make temporary connections to the NuASIC (Fig.~\ref{f:pins}). Each micro-pogo probe compresses independently to accommodate the variance in solder ball height and curvature. A micro-pogo probe guide encapsulates the spring structure of the probe, fixing the alignment to match the NuASIC pitch requirement while also limiting transverse deflection of the probes during vertical compression. The micro-pogo probes selected for the ATS were designed and produced by AlphaTest Corp., which specializes in fine-pitch test sockets. The micro-pogo probes have crowned 140 $\mu$m diameter heads that cup directly onto the NuASIC solder balls, while the flat 203 $\mu$m diameter ends of the pins are mounted onto pads on the ATS' readout board. Long 5.1 mm micro-pogo probes were selected to offset the TSV NuASIC from the surface of the ATS test board, allowing a gap between ASIC and test board for edge-on monitoring of alignment and compression with USB microscopes (Fig.~\ref{f:ats}b). Each micro-pogo spring can maximally travel 700 $\mu$m but the probe guide limits compression to 350 $\mu$m to preserve spring health and limit force on the NuASIC. In addition to an array of 87 micro-pogo probes for the TSV NuASIC readout, two arrays of 6\,$\times$\,6 probes make contact with dummy pads at the opposite edge of the TSV NuASIC. This distributes the compression force across the TSV NuASIC surface preventing the device from shifting. The variety of micro-pogo probes available from AlphaTest Corp. has also allowed the construction of a secondary micro-pogo test socket with pointed pogo-probes. This allows for testing NuASICs without solder balls in the event another integration mechanism is chosen instead of flip-chip bonding.

The ATS is a mechanical fixture designed to control the alignment and contact force of the micro-pogo probe socket with the TSV NuASIC under test (Fig.~\ref{f:ats}a). The TSV NuASIC is placed flat on a electrostatically dissipative (Semitron ESD-225) surface. An L-bracket is machined into the surface which assists with TSV NuASIC alignment. Additionally, the placement of the TSV NuASIC can be secured by a vacuum chuck located below the ESD-225 plate and routed through holes to the ASIC. In practice, the use of vacuum to secure the TSV NuASIC has not been required. The alignment of the micro-pogo probe test socket is adjusted relative to the TSV NuASIC's position by a micrometer-controlled 2D horizontal translation stage and a rotational stage. To guide and confirm alignment, two USB microscopes were used to allow us to observe the solder ball and micro-pogo probes edge-on as the test socket is lowered by a vertical micrometer stage. One microscope confirms alignment of the 87 solder balls while the second is used along the adjacent edge to adjust the micro-pogo probes relative to the two offset solder ball columns and dummy pads. Top-down USB endoscopes are incorporated into the design to align the ATS with fiducial points on the surface of the TSV NuASIC. These endoscopes typically are only used when solder balls are not present on the TSV NuASIC surface which makes edge-on alignment confirmation difficult. The TSV NuASIC test procedure is sensitive to light and other electromagnetic interference, so exposure is prevented by a light-tight grounded aluminum cover that is put in place after TSV NuASIC alignment and compression. Once the NuASIC is in contact with the micro-pogo pin test socket and has been covered, testing is controlled through the ATS test board.

\begin{figure}[tbh!]
   \centering
\includegraphics[width=6.0in]{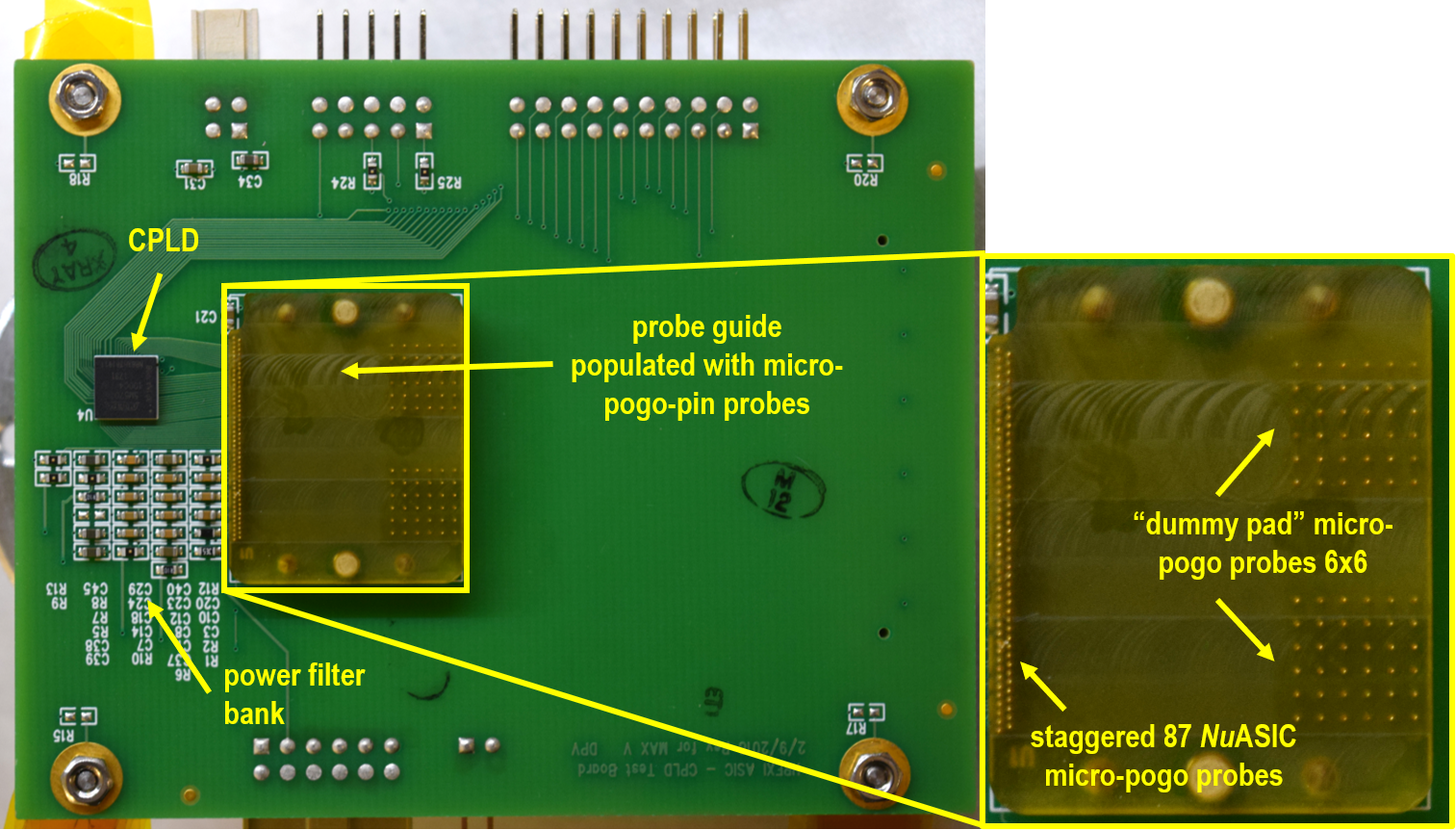}
\caption{NuASIC test board with micro-pogo pin test socket. The CPLD handles low-level signaling while passing data to an off-board FPGA. Power filter-networks on board support both the ASIC and CPLD. The micro-pogo probe guide includes a staggered line of 87 probes for ASIC operation and two 6\,$\times$\,6 grids of probes to distribute compression force on the device.}
    \label{f:testboard}
\end{figure}

The ATS test board is composed of the micro-pogo probe test structure, a Complex Programmable Logic Device (CPLD) and passives that form a low-pass power filter for both the CPLD and NuASIC (Fig.~\ref{f:testboard}). The CPLD handles low-level NuASIC signaling and passes user commands to the NuASIC from a commercial-off-the-shelf Field Programmable Gate Array (FPGA) that sits behind a power isolation barrier on a secondary board. Data from the NuASIC is passed through the CPLD off the test board to the FPGA where it is buffered. The FPGA handles the event data packaging for the final data product and passes both data and commands serially over USB with a user-controlled computer. Test pins on the ATS test board allow for confirmation of proper connection with the TSV NuASIC by confirming low resistivity measurements between the CPLD and TSV NuASIC ground planes.

On the ATS, the test board drives TSV NuASIC evaluation via an internal test pulser\cite{Harrison10}. During testing, the NuASIC test pulser injects charge into selectable contiguous arrays of the 1024 NuASIC pixels. This injected charge mimics the charge collection pathway of a NuASIC bonded to CZT when exposed to an X-ray event.  The current inducted by the injection of this charge is copied sequentially to a series of 16 capacitors on the NuASIC.  The collected charge on each capacitor is digitized in an on-chip digital-to-analog converter and read out by the detector control system.In the most stable testing configuration, 4\,$\times$\,4 pixel regions on the TSV NuASIC are stimulated sequentially with the test pulser for several minutes. All 1024 NuASIC pixels can be well-characterized within 30 minutes, and the data can be used to produce a report of individual pixel health status, resolution, and gain.

\section{Performance of NuASICs tested through the ATS} \label{s:performance}

\begin{figure*}[tbh!]
\centering
   \includegraphics[width=6.0in]{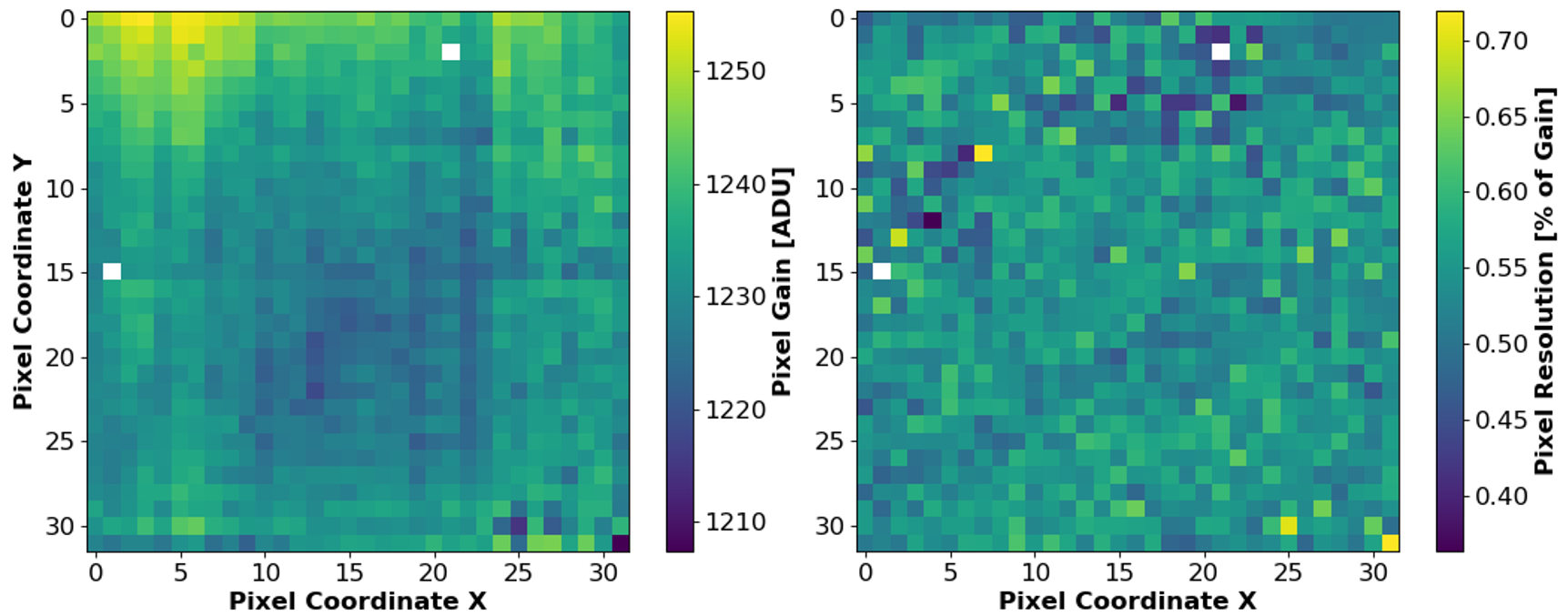}
\caption{\textbf{\textit{(Left}}) Gain pixel map of diced, unbonded TSV NuASIC. White pixels represent dead, unresponsive, or masked pixels. An ADU gain value of 1230 approximately corresponds to a 130keV pixel trigger. (\textbf{\textit{Right}}) Resolution map of diced TSV NuASIC. White pixels represent dead, unresponsive, or masked pixels. FWHM of each pixel gain reported as a \% of the channel's gain.}
    \label{f:2dbackside}
\end{figure*}

\begin{figure}[tbh!]
   \centering
\includegraphics[width=4.5in]{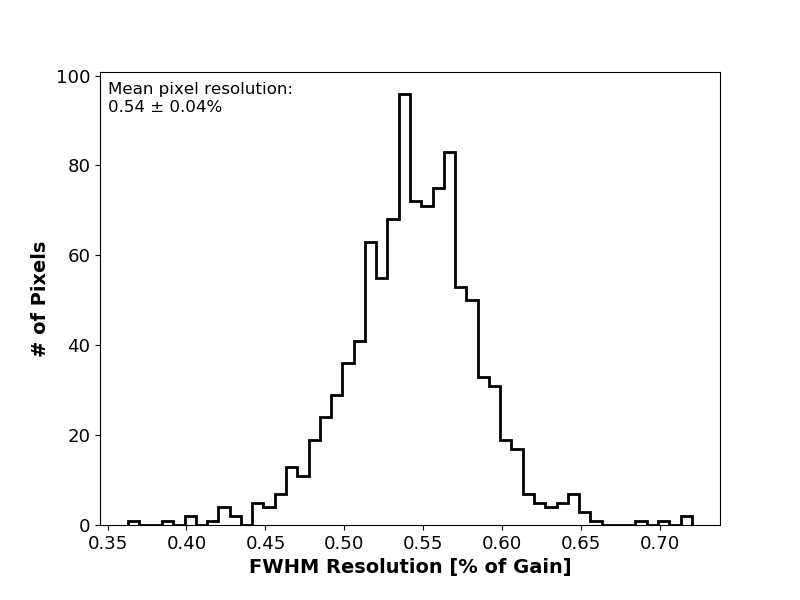}
\caption{Histogram of FWHM resolutions of the back-side \textit{blind} TSV NuASIC's 1024 (1022 active) pixel channels as a percentage of total gain. The mean resolution of the TSV NuASIC tested corresponds to 0.54\% (6.7 ADU).}
\label{f:histobackside}
\end{figure}

We have successfully used the ATS to test TSV NuASICs modified by both back-side \textit{blind} TSVs and front-side \textit{through} TSVs incorporated through via-last processes. Back-side TSVs are implemented from the bottom of the NuASIC by inserting a single 100 $\mu$m diameter by 300 $\mu$m deep tapered cylindrical TSV that touches down directly onto the back of each of the wire-bond pads on the top surface. Figure \ref{f:2dbackside} displays a full internal pulser scan of a back-side TSV NuASIC performed by sweeping the internal pulser across the 32\,$\times$\,32 NuASIC pixels in a 4\,$\times$\,4 grid and collecting data from the triggered channels. ``Hot'' (noisy) pixels or unresponsive pixels are filtered out from the image to determine the overall number of functioning pixel channels in the device, resulting in 1022 active pixels out of 1024 for the device under test. The scan also reports the mean analog-to-digital unit (ADU) value of each pixel channel with an ADU value of 1230 corresponding approximately to a 130 keV X-ray event. The resolution of each pixel as a full width at half maximum (FWHM) percentage of the ADU is also mapped in Figure \ref{f:2dbackside}. A histogram of FWHM resolutions of the 1022 active pixels from the back-side TSV NuASIC is found in Figure \ref{f:histobackside}. We find an improved average resolution (0.54\% FWHM) in comparison to data reported in Hong et al.\,(2021)\cite{Hong21} from NuASIC internal pulser testing with a traditional probe station on the wafer level prior to dicing. This indicates that the testing performance of the ATS is stable and a capable replacement to traditional probe card testing.

\begin{figure}[tbh!]
   \centering
\includegraphics[width=5.0in]{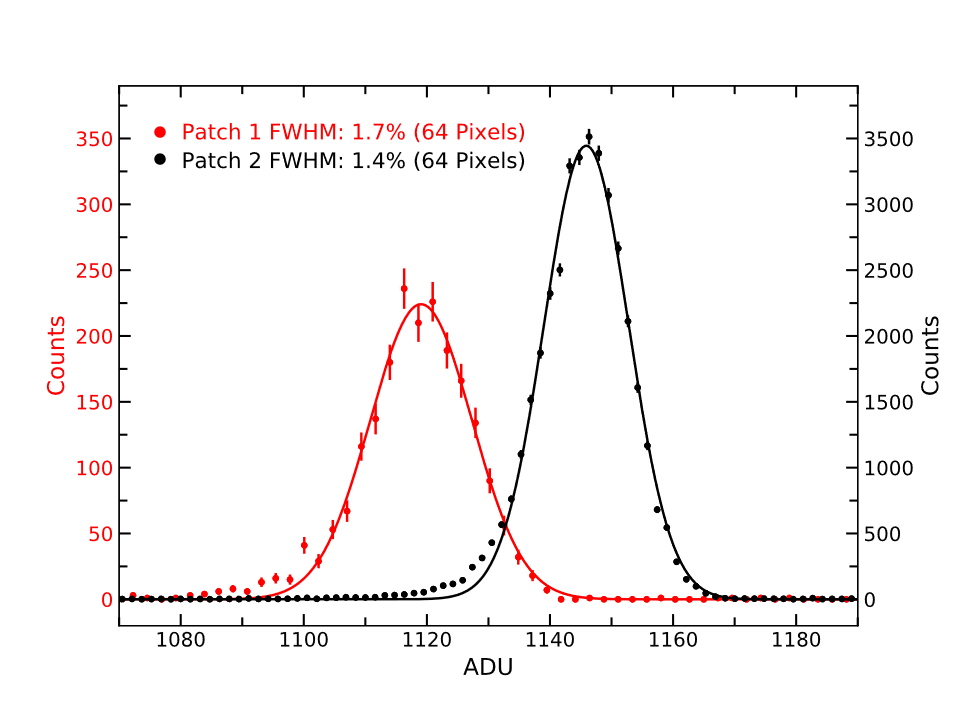}
\caption{Pulser spectra of two 8\,$\times$\,8 pixel regions on opposite corners of a front side {\it through}-TSV NuASIC tested with the ATS. Patch 1 has a 20 ADU FWHM while pixel patch 2 has a 14 ADU FWHM.}
    \label{f:histofrontside}
\end{figure}

Challenges with implementing the back-side TSVs successfully\cite{Hong21} led to the exploration of a front-side \textit{through} TSV design with Micross Advanced Interconnect Technology\cite{Ovental21}. In this process, three 20 $\mu$m diameter TSVs are etched adjacent to each of the wire bond pads, then revealed on the backside of the NuASICs through a polishing and etching process before backside traces are deposited. The first run of the front-side TSV NuASICs tested with the ATS produced a functioning device yield of approximately 30\%. While these devices returned high resolution pulser spectra at first, the functionality of the front-side TSV NuASICs degraded quickly with increasing current draws, becoming unresponsive in approximately 10 minutes of operation time\cite{Hong21}. While this prevented a full ATS scan of a front-side TSV NuASIC, we performed limited testing on regions of pixels with the spectral resolution of the internal pulser reported in Figure \ref{f:histofrontside}. We tested two 8\,$\times$\,8 patches of pixels on opposite corners of the front-side TSV NuASIC and between tests observed a shift in peak ADU value which may be a result of the device's changing performance. Both pixel patches' FWHM spectral resolution remained less than 2\% of the total ADU value, similar to wafer-level testing performed and reported in Hong et al.\,(2021)\cite{Hong21}. The low yield in front-side TSV NuASICs from the first experimental run most likely results from an insufficient amount of insulation or isolation around the TSVs and is being further explored. 

TSV NuASICs were inspected before and after testing with the ATS to determine if the compression forces from using an array of micro-pogo probes could damage the devices. Of particular concern were the front-side \textit{through} TSV NuASICs, as the TSV implementation process results in additional thinning of the devices to expose the vias. However, no damage to the TSV NuASICs silicon was observed, while repeated compression on both TSV NuASICs pads and solder balls only resulted in slight indentations. We previously confirmed while operating the micro-pogo pin test socket and ATS that a limited probe compression of 50-100 $\mu$m is necessary to achieve sufficient probe contact with the TSV NuASIC, reducing the compression force necessary for testing\cite{Violette18}.


\section{Summary and Future Development} \label{s:futureS}

We have successfully designed, constructed and tested the functionality of a die-level test stand that will allow for the rapid screening of individual TSV NuASICs. This assures that low-noise NuASICs can be selected for integration into HREXI's closely tiled CZT/TSV NuASIC 16\,$\times$\,16 imaging detector array for a wide-field coded aperture X-ray telescope at lower cost, risk and complexity than previous wire bonds designs. The ATS allows for rapid pixel-to-pixel channel calibration of the 32\,$\times$\,32 pixels of each TSV NuASIC. The ATS will be valuable as the development of the front-side \textit{through} TSV NuASIC process continues, offering rapid testing of NuASIC die to determine wafer yield and functionality. Additionally, a secondary operating mode available on the NuASIC allows the detector to achieve greater spectral resolution at low (3-5 keV) X-ray energies at the cost of leakage current sensitivity. This secondary operating mode, currently used on the \textit{NuSTAR} mission, has tunable parameters that must be adjusted for improved operation prior to launch. The ATS will serve as an integral testbench for the parameter adjustments necessary to operate the TSV NuASICs in this mode.

HREXI will require the evaluation of more CZT/NuASIC detectors than any previously launched instrument. While the ATS successfully minimizes the necessary human resources for testing via rapid alignment capabilities, the NuASIC testing duration is dominated by the pixel pulser scan which can take up to thirty minutes to achieve reasonable spectral resolution statistics. For HREXI's proposed 256 unit CZT/TSV NuASIC detector array, the testing of all individual NuASICs will take nearly a week. To accommodate this, the ATS can be duplicated to allow for parallel testing of multiple NuASICs.

The development of a future SmallSat Constellation capable of full-sky hard X-ray monitoring will further expand detector testing requirements. Continued ATS development will fine-tune micro-pogo probe test fixtures with smaller probe diameters and finer pitch for testing successor ASICs with 2$\times$ finer spatial resolution (64\,$\times$\,64 pixels). These advances in ATS development will promote generalization of the test stand to future, higher resolution ASICs that will be used in next-generation instrument designs.

\section{Acknowledgements}

We would like to thank AlphaTest Corp. for their cooperation in helping design and manufacture the micro-pogo probes and probe socket and mounting it onto our ATS test board. We would like to thank Micross AIT for their continued support on developing front-side \textit{through} NuASICs. DPV is grateful to MA and KV for useful advice. This work was supported by NASA APRA grant NNX17AE62G. DPV is supported by the NASA FINESST Fellowship 80NSSC20K1537.



\bibliography{references} 

\begin{thebibliography}{10}

\bibitem{Hong09}
J.~{Hong}, B.~{Allen}, J.~{Grindlay}, {\em et~al.}, ``{Building large area CZT
  imaging detectors for a wide-field hard X-ray
  telescope{\textemdash}ProtoEXIST1},'' {\em Nuclear Instruments and Methods in
  Physics Research A} {\bf 605}, 364--373  (2009).

\bibitem{Allen11}
B.~Allen, J.~Hong, J.~Grindlay, {\em et~al.}, ``{Development of the ProtoEXIST2
  advanced CZT detector plane},'' {\em Proc. IEEE Nuclear Science Symp. and
  Medical Imaging Conf.} , 4470--4480  (2011).

\bibitem{Hong13}
J.~{Hong}, B.~{Allen}, J.~{Grindlay}, {\em et~al.}, ``{Tiled Array of Pixelated
  CZT Imaging Detectors for ProtoEXIST2 and MIRAX-HXI},'' {\em IEEE
  Transactions on Nuclear Science} {\bf 60}, 4610--4617  (2013).

\bibitem{Hong17}
J.~{Hong}, B.~{Allen}, J.~{Grindlay}, {\em et~al.}, ``{Imaging Analysis of the
  Hard X-Ray Telescope ProtoEXIST2 and New Techniques for High-resolution
  Coded-aperture Telescopes},'' {\em AJ} {\bf 153}, 11  (2017).

\bibitem{Harrison13}
F.~A. {Harrison}, W.~W. {Craig}, F.~E. {Christensen}, {\em et~al.}, ``{The
  Nuclear Spectroscopic Telescope Array (NuSTAR) High-energy X-Ray Mission},''
  {\em ApJ} {\bf 770}, 103  (2013).

\bibitem{Harrison10}
F.~A. Harrison, W.~R. Cook, H.~Miyasaka, {\em et~al.}, {\em Semiconductor
  Radiation Detection Systems}, CRC Press, Boca Raton, FL  (2010).

\bibitem{Hong21}
J.~{Hong}, J.~{Grindlay}, B.~{Allen}, {\em et~al.}, ``{Proof of concept for
  through silicon vias in application-specific integrated circuits for hard
  x-ray imaging detectors},'' {\em Journal of Astronomical Telescopes,
  Instruments, and Systems} {\bf 7}, 026001  (2021).

\bibitem{Hong17b}
J.~Hong, B.~Allen, J.~Grindlay, {\em et~al.}, ``Through-{S}ilicon-{V}ias
  ({TSV}s) for {3D} readout of {ASIC} for nearly gapless {CdZnTe} detector
  arrays,'' {\em Proc. SPIE} {\bf 10392}  (2017).

\bibitem{Violette18}
D.~{Violette}, B.~{Allen}, J.~S. {Hong}, {\em et~al.}, ``{Efficient validation
  testing of Through-Silicon-Via (TSV) ASICs for CZT x-ray detectors},'' in
  {\em Hard X-Ray, Gamma-Ray, and Neutron Detector Physics XX},  {\em Society
  of Photo-Optical Instrumentation Engineers (SPIE) Conference Series} {\bf
  10762}, 107620S  (2018).

\bibitem{Ovental21}
J.~{Ovental}, D.~{Malta}, D.~{Bordelon}, {\em et~al.}, ``{TSV-Last Integration
  to Replace ASIC Wire Bonds in the Assembly of X-Ray Detector Arrays},'' {\em
  Proceedings of IEEE 71st Electronic Components and Technology Conference} ,
  170--177  (2021).

\end{thebibliography}
\bibliographystyle{spiejour}


\section*{Biographies}

{\bf Daniel Violette} is a graduate student at Harvard University working with the High-Resolution Energetic X-ray Imager (HREXI) team, with interests in high-energy time domain astrophysics and instrumentation. Daniel is supported by the Future Investigators in NASA Earth and Space Science and Technology (FINESST) Fellowship to further develop HREXI detector sensitivity at low energy thresholds.

{\bf Branden Allen} received his Ph.D.~degree in physics from U.C. Irvine in 2007 and is currently a Senior Research Scientist at Harvard University with over 20 years of experience in the development, deployment and operation of ground- and space-based telescopes for high-energy X/$\gamma$-ray astronomy and planetary science.  His current research is focused on the development and deployment of next generation detector systems and telescopes to probe high energy astrophysical phenomena and for future planetary exploration.  

{\bf Jaesub Hong} is a Senior Research Scientist at Harvard University. He has nearly 20 years of experience in development of X-ray telescopes for high energy astrophysics and planetary science.  His current focus is  the development of advanced hard X-ray detectors for next generation wide-field hard X-ray telescopes for time domain astrophysics and the miniature lightweight X-ray optics for planetary science.  He received a Ph.D.~degree in Physics from Columbia University. He has (co)authored over 40 publications. 

{\bf Hiromasa Miyasaka} is a staff scientist at California Institute of Technology. He received a Ph.D.~in Physics (2000) from Saitama University in Japan. He has over 20 years of experience in development of particles and X-ray detectors for the cosmic ray and high-energy astrophysics. Since 2006, his work has focused on CdZnTe and CdTe detectors and readout ASIC development. He is one of the primary detector scientists for the \textit{NuSTAR} mission.

{\bf Jonathan Grindlay} is the Robert Treat Paine Professor of Astronomy at Harvard. He received his BA in Physics from Dartmouth (1966) and PhD in Astrophysics from Harvard  (1971). He joined the Faculty in 1976 and  Chaired the Department in 1985-91 and 2001-03. His primary interest is black hole time variability, accretion physics, accreting black hole (both stellar and supermassive) populations and formation as measured with wide-field coded aperture imaging X-ray telescopes (ultimately full-sky) and optical/IR imaging/spectroscopy. He has over 434 refereed Journal papers.

\newpage

\section*{Figures}

\end{document}